\documentclass[%
 reprint,
 amsmath,amssymb,
 twocolumn,
 aps,
 prx,
floatfix,
longbibliography,
superscriptaddress
]{revtex4-2}

\makeatletter
\def\@bibdataout@aps{%
 \immediate\write\@bibdataout{%
  @CONTROL{%
   apsrev41Control,author="08",editor="1",pages="0",title="0",year="1"%
  }%
 }%
 \if@filesw
  \immediate\write\@auxout{\string\citation{apsrev41Control}}%
 \fi
}%
\makeatother 

\usepackage{graphicx}
\usepackage{dcolumn}
\usepackage{bm}
\usepackage{layouts}
\usepackage[colorlinks,linkcolor=blue,anchorcolor=blue,citecolor=blue,urlcolor=blue]{hyperref}
\usepackage{float}
\usepackage{makecell}
\usepackage{verbatim}

\begin{document}

\preprint{APS/123-QED}

\title{Quantum Non-Demolition Measurement on the Spin Precession of Laser-Trapped $^{171}$Yb Atoms}

\affiliation{CAS Center for Excellence in Quantum Information and Quantum Physics, School of Physical Sciences, University of Science and Technology of China, Hefei 230026, China}
\affiliation{Hefei National Laboratory, University of Science and Technology of China, Hefei 230088, China}
\affiliation{CAS Key Laboratory of Quantum Information,\\ University of Science and
Technology of China, Hefei 230026, China.}

\author{Y. A. Yang}
\thanks{These authors contributed equally to this work.}
\affiliation{CAS Center for Excellence in Quantum Information and Quantum Physics, School of Physical Sciences, University of Science and Technology of China, Hefei 230026, China}

\author{T. A. Zheng}
\thanks{These authors contributed equally to this work.}
\affiliation{CAS Center for Excellence in Quantum Information and Quantum Physics, School of Physical Sciences, University of Science and Technology of China, Hefei 230026, China}

\author{S.-Z. Wang}
\affiliation{CAS Center for Excellence in Quantum Information and Quantum Physics, School of Physical Sciences, University of Science and Technology of China, Hefei 230026, China}
\author{W.-K. Hu}
\affiliation{CAS Center for Excellence in Quantum Information and Quantum Physics, School of Physical Sciences, University of Science and Technology of China, Hefei 230026, China}
\author{Chang-Ling Zou}
\affiliation{CAS Center for Excellence in Quantum Information and Quantum Physics, School of Physical Sciences, University of Science and Technology of China, Hefei 230026, China}
\affiliation{Hefei National Laboratory, University of Science and Technology of China, Hefei 230088, China}
\affiliation{CAS Key Laboratory of Quantum Information,\\ University of Science and
Technology of China, Hefei 230026, China.}
\author{T. Xia}%
\thanks{Corresponding author: txia1@ustc.edu.cn}
\affiliation{CAS Center for Excellence in Quantum Information and Quantum Physics, School of Physical Sciences, University of Science and Technology of China, Hefei 230026, China}
\author{Z.-T. Lu}%
\thanks{Corresponding author: ztlu@ustc.edu.cn}
\affiliation{CAS Center for Excellence in Quantum Information and Quantum Physics, School of Physical Sciences, University of Science and Technology of China, Hefei 230026, China}
\affiliation{Hefei National Laboratory, University of Science and Technology of China, Hefei 230088, China}

\date{\today}

\begin{abstract}
Quantum non-demolition (QND) measurement enhances the detection efficiency and measurement fidelity, and is highly desired for its applications in precision measurements and quantum information processing. We propose and demonstrate a QND measurement scheme for the spin states of laser-trapped atoms. On $^{171}$Yb atoms held in an optical dipole trap, a transition that is simultaneously cycling, spin-state selective, and spin-state preserving is created by introducing a circularly polarized beam of control laser to optically dress the spin states in the excited level, while leaving the spin states in the ground level unperturbed. We measure the phase of spin precession of $5\times10^{4}$ atoms in a bias magnetic field of 20 mG. This QND approach reduces the optical absorption detection noise by $\sim$19 dB, to a level of 2.3 dB below the atomic quantum projection noise. In addition to providing a general approach for efficient spin-state readout, this all-optical technique allows quick switching and real-time programming for quantum sensing and quantum information processing.

\end{abstract}

\maketitle


\section{\label{introduction}Introduction}

The nuclear and electronic spin states of atoms, with the advantage of having long coherence times, are of great importance in precision measurement and quantum information experiments. Measurements on atomic spin states are performed to realize magnetometers~\citep{budker2007optical,Limes2018,Allred2002},
optical clocks~\citep{Ye2008,Campbell2017,Ludlow2015,Schioppo2017},
and in experiments that search for permanent electric dipole moment~\citep{Safronova18,Chupp2019a,ParDieKal15,Cairncross2017,Andreev2018,Lim2018,Graner2016b,Sachdeva2019a}.
As an information carrier, the spin state is also widely used in quantum information processing~\citep{GorReyDal09,Noguchi2011,Komar2016,Yang2020,Jenkins2021a,Ma2022,Chen2022,Barnes2022} and quantum simulation~\citep{Bernien2017,Dai2016}. For these applications, it is often crucial to measure the population in each spin state with a high efficiency in order to reduce the statistical errors in precision measurements and to enhance the readout fidelities in quantum information experiments.

The populations in spin states can be determined either dispersively  by measuring the state-dependent phase shift of an off-resonant laser beam, or dissipatively by measuring the absorption of or the fluorescence induced by near-resonant light. In the recent quantum computing experiments, it is challenging to achieve high fidelity of spin selectivity in the qubit-state readout~\citep{Ma2022,Jenkins2021a,Barnes2022}. The detection efficiency is often limited by measurement-induced spin flipping, upon which the quantum state is demolished prematurely. For example,
probing the states of a spin-1/2 system on non-cycling transitions, as shown in Figs.~\ref{transition}(a) and ~\ref{transition}(b), induces a spin flip with just a few excitation–emission cycles~\citep{ParDieKal15,GorReyDal09}. Alternatively, in a quantum non-demolition (QND) measurement~\citep{Braginsky1996,Poizat2007a,Nogues1999}, the state is preserved under repeated excitation–emission cycles, thus the signal-to-noise ratio of state detection can be greatly enhanced. For this reason, QND measurements have attracted increasing interests and have been successfully demonstrated on various systems achieving improved measurement fidelity in quantum information processing~\citep{Xue2020,Nakajima2019,Lupascu2007} and higher precision in quantum measurements~\citep{Pezze2018,Bao2020,Bowden20}. 

Several QND strategies for spin-state detection have been demonstrated. For example, a magnetic field can be applied to lift the degeneracy among transitions of different magnetic sublevels~\citep{Braverman2019}, under the condition that the induced Zeeman splittings are much larger than the transition linewidth. While this method works efficiently, the required magnetic field and its on-and-off switching can disturb the spin states, resulting in decoherence and loss of sensitivity. Another strategy, more suitable for solid-state systems, is to modify the density of states and induce a state-dependent spontaneous relaxation rate via the Purcell effect~\citep{PhysRev.69.674}. Recently, such a scheme is implemented for single rare-earth ions embedded in a nano-photonic cavity, boosting the transition cyclicity by several orders of magnitude~\citep{Chen592}.

In this work, we present a theoretical analysis and an experimental demonstration of a QND approach to probe spin states and measure the phase of spin precession via optical excitation. By applying an ancillary control laser to shift the excited states via the ac-Stark effect, while leaving the spin states in the ground level unperturbed, the chosen optical transition can simultaneously become cycling, spin-selective, and spin-preserving. AC Stark shift has been successfully employed in many applications, including state-selective manipulation of atomic internal states~\cite{Eto,DeLeseleuc2017,Park2001}, site-selective addressing in atom array~\cite{Weitenberg2011,Labuhn2014,Xia2015,Wang2015,Wang2016}, and narrow-line Sisyphus cooling~\cite{Taieb1994,Ivanov2011,Cooper2018b,Covey2019}. Our approach is demonstrated on $^{171}$Yb atoms in an optical dipole trap (ODT) whose wavelength satisfies the magic condition for the probe transition~\cite{Zheng2020}. QND measurements are performed on spin precession, demonstrating a reduction in optical noise by $\sim$19 dB. This all-optical approach of QND measurement avoids the need to switch and shield any control magnetic fields. Its principle can be applied to many different atomic systems and is compatible with general cold-atom experiments in precision measurements and quantum information science. This novel method was introduced in Ref.\citep{Zheng2022}, where it was used in a measurement of the electric dipole moment of $^{171}$Yb.

\section{\label{principle}Principle}
\begin{figure}
	\includegraphics[width=3.4in]{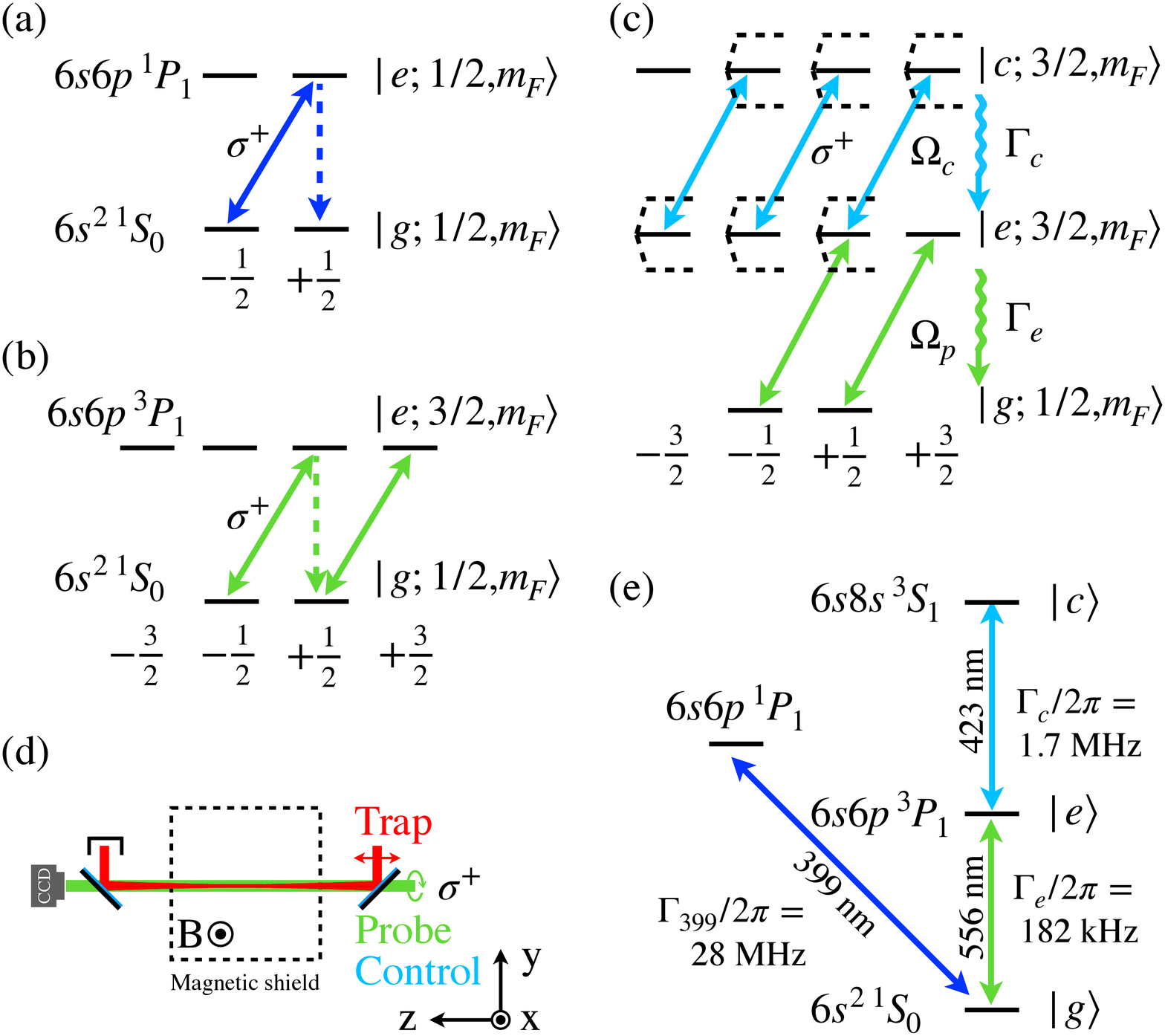}
  \caption
  {(a) The $F=1/2\leftrightarrow F=1/2$ case for detection of nuclear spin states in a non-QND method. An atom in the dark state $|g; 1/2, +1/2\rangle$ does not scatter photons; an atom in $|g; 1/2, -1/2\rangle$ scatters on average 3 photons before a spin flip occurs. 
    (b) The $F=1/2\leftrightarrow F=3/2$ case for detection of nuclear spin states in a non-QND method. An atom in $|g; 1/2, +1/2\rangle$ can be probed on a cycling transition; an atom in $|g; 1/2, -1/2\rangle$ scatters on average 1.5 photons before a spin flip occurs. 
    (c) The proposed QND approach for detection of nuclear spin states. The optical dressing of the $|e\rangle$ state is by the control beam. In this condition, the $|g; 1/2, -1/2\rangle$ $\leftrightarrow$ $|e; 3/2, +1/2\rangle$ transition is optically shifted and suppressed, while $|g; 1/2, +1/2\rangle$ $\leftrightarrow$ $|e; 3/2, +3/2\rangle$ remains unaffected. The spin state can be probed without risking spin flips. 
    (d) Layout of the setup. The atoms are trapped by an optical dipole trap (ODT) in a vacuum chamber and $\mu$-metal shields (dashed rectangle). A $\cos \theta$ coil is used to generate a uniform bias magnetic field of 20 mG in the $x$ direction. The ODT, polarization beam, probe beam, and control beam are all combined and directed in the $z$ direction. ODT is linearly polarized in the $y$ direction; the polarization, probe, and control beams have the same $\sigma^{+}$ circular polarization.
    (e) Energy levels and transitions of  Yb. In the QND approach, the probe is on the intercombination transition at  556  nm; the control transition is at 423 nm. The fast singlet-to-singlet transition at 399 nm is used for optical pumping to produce spin polarization, and for detecting spin states in the non-QND method.}
  
\label{transition}
\end{figure}

The principle of optical pumping and spin-state detection is often explained with the simple case of $F=1/2\leftrightarrow F=1/2$ [Fig.~\ref{transition}(a)]. Throughout this paper, the quantization direction is chosen to be along the common $\hat{k}$ vector of the polarization beam, probe beam and control beam. A laser beam of resonant frequency and $\sigma^{+}$ polarization excites the $m_{F}=-1/2$ state in the ground level, but not the $m_{F}=+1/2$ “dark state”. In this case, an atom in $|g;1/2,-1/2\rangle$ absorbs and emits on average only three photons before dropping into $|g;1/2,+1/2\rangle$, thus limiting the fidelity of state detection. For the case of a different transition, $F=1/2\leftrightarrow F=3/2$ [Fig.~\ref{transition}(b)], even though an atom in $|g;1/2,+1/2\rangle$ can be probed repeatedly on the $|g;1/2,+1/2\rangle$ $\leftrightarrow$ $|e;3/2,+3/2\rangle$ cycling transition, an atom in $|g;1/2,-1/2\rangle$ absorbs and emits on average only 1.5 photons before a spin flip occurs. For a QND measurement, we need a transition that is simultaneously cycling, spin selective and spin preserving. 

We propose a QND measurement scheme based on the optical dressing effect. Consider a ladder-type atomic system with three levels: the ground level $|g\rangle$, the excited level $|e\rangle$, and the excited level $|c\rangle$ [Fig.~\ref{transition}(c)] ($|c\rangle$ can be either higher or lower than $|e\rangle$ in energy). Their angular momenta are 1/2, 3/2, and 3/2, and spontaneous decay rates are 0, $\Gamma_{e}$, and $\Gamma_{c}$, respectively. The control beam, on resonance with the $|e\rangle\leftrightarrow|c\rangle$ transition, dresses the $|e\rangle$ state with Rabi frequency $\Omega_{c}$. The optical dressing is on all Zeeman states $|e; 3/2, m_F\rangle$, except for the stretched state $|e; 3/2, +3/2\rangle$. The dressed Zeeman states are shifted by $\pm\Omega_{c}/2$ to form Aulter-Towns doublets~\citep{Khan2016a} and the stretched state is protected by the angular momentum selection rules. Such a difference among the Zeeman states is essential for spin-selective detection of the spin state.

To probe the nuclear spin state, the probe beam resonantly drives the $|g\rangle\leftrightarrow|e\rangle$ transition with Rabi frequency $\Omega_{p}$. In the condition where $\Omega_{p}\ll\Omega_{c},\Gamma_{e}$, the optical transition strength to the dressed Zeeman states is reduced by a factor of $\sim \Omega_{c}^{2} / (\Gamma_{e}\Gamma_{c})$. The $|g; 1/2, -1/2\rangle$ state then becomes a “dark state” and the rate of spin flip is reduced by the same factor. With modest intensity of the control beam, the reduction factor can be on the order of $10^{3}$. The atoms in $|g; 1/2, +1/2\rangle$ can be excited to the unaffected stretched state repeatedly, and optical detection of the nuclear spin state with cycling transition is realized. It is important to note that the control beam can be switched on and off at a rate much faster than the spin precession rate, and does not affect the spin states in the ground level.


\section{\label{experiment}Experimental setup}
We have implemented the QND measurement on the spin states of $^{171}$Yb (I = 1/2) atoms in the ground level. The three levels $6s^2\,{}^1S_0\ (F=1/2)$, $6s6p\,{}^3P_1\ (F=3/2)$ and $6s8s\,{}^3S_1\ (F=3/2)$ form a ladder system as shown in Fig.~\ref{transition}(e). The QND measurement employs the following laser beams and transitions: The probe beam at 556 nm, tuned to the resonance of $6s^2\,{}^1S_0\leftrightarrow 6s6p\,{}^3P_1$ ($\Gamma_{e}/2\pi=182$ kHz), is supplied by a frequency-doubled diode laser; the control beam at 423 nm, tuned to the resonance of $6s6p\,{}^3P_1\leftrightarrow 6s8s\,{}^3S_1$($\Gamma_{c}=2\pi\times1.7$~MHz), is supplied by a frequency-doubled Ti:Sapphire laser; the polarization beam at 399 nm, tuned to the resonance of $6s^2\,{}^1S_0\leftrightarrow 6s6p\,{}^1P_1$($\Gamma_{399}/2\pi=28$ MHz), is supplied by a frequency-doubled diode laser. The probe beam, the control beam, and the polarization beam all have the same circular polarization (e.g., $\sigma^{+}$), and all co-propagate with the stationary ODT beam along the $z$ direction [Fig.~\ref{transition}(d)]. The absorption of the probe beam by the trapped atoms is imaged onto a CMOS camera.

To prepare the atomic ensemble, $^{171}$Yb atoms are loaded into a two-stage magneto-optical trap (MOT): The first-stage MOT is operated on the strong transition (same as the polarization transition) to efficiently capture the atoms from a Zeeman slower; the  second-stage  MOT  on  the narrow-linewidth intercombination transition (same as the probe transition) cools the atoms to 20 $\mu$K. The cold atoms are then handed over to a movable ODT. More details of the apparatus are given in Ref~\cite{Zheng2020}. The atoms are carried into a neighboring science chamber by translating the focal point of the movable ODT along the $y$ direction [Fig.~\ref{transition}(d)] and, finally, handed over to a stationary ODT pointed in the $z$ direction. The two ODTs are provided by two separate fiber lasers. We prepare $10^{3}-10^{4}$ $^{171}$Yb atoms in the ODT for measurements. A $\cos \theta$ coil~\cite{Everett1966} inside magnetic shields generates a uniform B field ($\sim$ 20 mG) in the $x$ direction to drive spin precession.

The stationary ODT in the science chamber has a waist of $50\ \mathrm{\mu m}$ and a Rayleigh length of $\sim$ 4 mm, is linearly  polarized in the $y$ direction, and has a power of 35 W and a wavelength of 1035.84 nm. This wavelength meets the magic condition for the probe transition so that the probe remains effective despite of the deep trapping potential of 200 $\mu$K~\cite{Zheng2020}. The probe linewidth is measured to be $\sim$ 400 kHz, reflecting Doppler broadening at 100 $\mu$K. The vector light shift of the spin states introduced by the linearly polarized ODT beam is negligible (\textless\ 1 mHz). The control beam is focused on the atoms with a beam waist of about 300 $\mu$m. The parameters for the control beam is determined by measuring the induced light shifts of the probe transition (Appendix \ref{speccontrol}). At the control beam power of 40 mW ($\Omega_{c}\sim 2\pi\times40$ MHz), the spin-flip rate in the ground level is reduced by a factor of $\Omega_{c}^{2}/(\Gamma_{e}\Gamma_{c})\sim 10^{3}$. Since the control beam is far detuned from any transitions that connect to the ground level, its effects on spin precession are negligible: the scalar light shift of $|g\rangle$ induced by the control beam is at $\sim$ kHz, much less than $\Gamma_{e}$; the vector light shift is at $\sim$ mHz, much less than the precession frequency of 15 Hz.

\section{\label{larmorprecession}Phase measurement of spin precession}

\begin{figure*}[t]
	\includegraphics[width=7.2in]{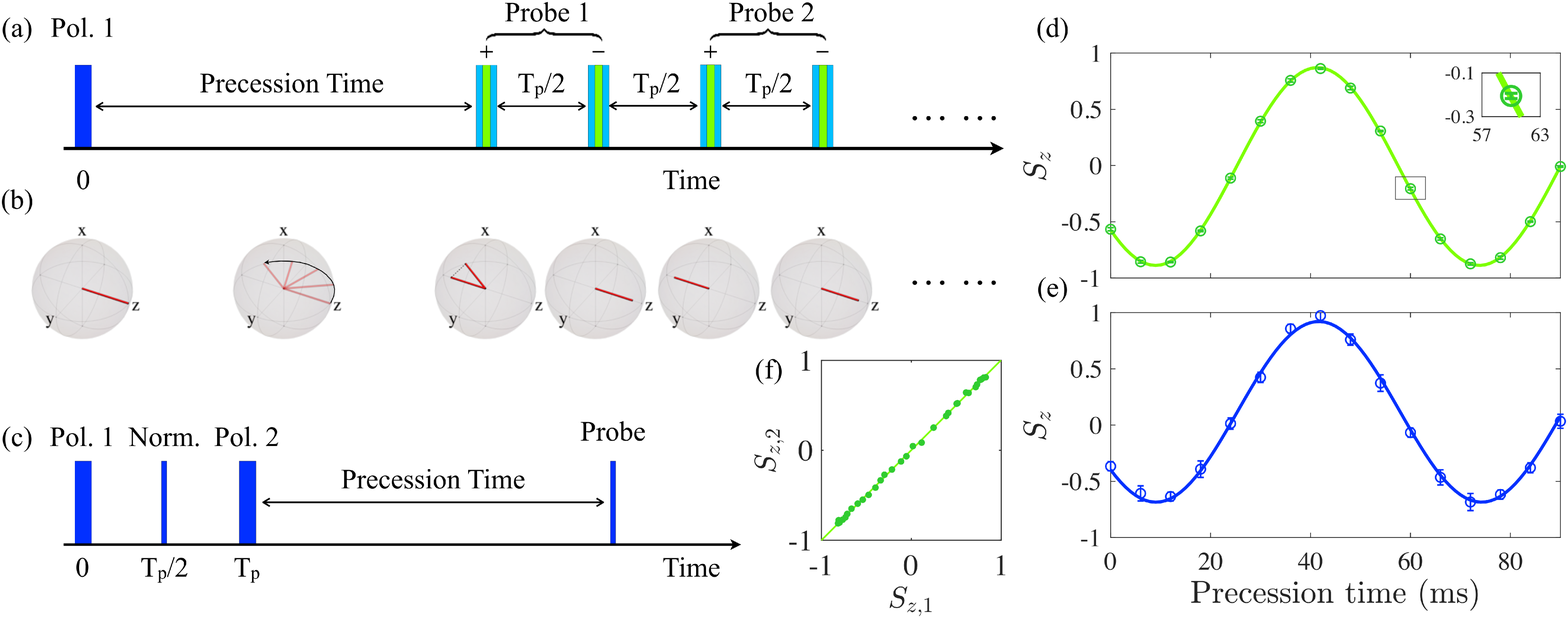}
  \caption
   {(a) Timing sequence of the QND measurement. Atoms are initially polarized by a 399 nm pulse (“Pol. 1”, $\color[rgb]{0.015,0.2,1}\blacksquare$). After a given precession time, two 556 nm probe pulses (“Probe 1”, $\color[rgb]{0.498,1,0}\blacksquare$), separated by $T_{p}/2$, measure the populations on $m_{F}=\pm1/2$ states successively. Each probe pulse is overlapped with a 423 nm control pulse ($\color[rgb]{0,0.753,1}\blacksquare$). The probe pulses can be repeated multiple times. (b) The evolution of the atomic spin Bloch vector under a QND measurement. (c) Timing sequence of a non-QND optical pumping measurement. A normalization pulse (“Norm.”, $\color[rgb]{0.015,0.2,1}\blacksquare$) is needed to measure the total population. The spin state population can only be measured once with only a few excitation-emission cycles. (d) Spin precession in the QND measurement. The initial spin polarization is about 90\%. The precession time is chosen to be 1s. (e) Spin precession measured with the non-QND method, with other parameters identical to those of (d). (f) $S_{z,1}$ is highly correlated with $S_{z,2}$. The reduced $\chi^{2}=1.03$ for the proportional fit (solid line).}
\label{sequence}
\end{figure*}

We demonstrate the advantage of the QND approach with phase measurements of the spin precession of $^{171}$Yb atoms. The timing sequence for the QND measurement is shown in Fig.~\ref{sequence}(a). Initial spin polarization is produced by a 2 ms pulse of the polarization beam (“Pol. 1”, $I/I_{s}=3\times10^{-4}$). The spin polarized atoms precess about the bias magnetic field ($\sim$ 20 mG) at a Larmor frequency of $\sim$ 15 Hz. After a given precession time, chosen to be 1 s in this study, an overlapping pulse of both the probe beam (0.4 ms, $I/I_{s}=0.25$) and control beam, named “Probe 1+”, is applied for a spin projection measurement. The population in $|g; 1/2, +1/2\rangle$ ($\rho_{+}$) is measured, while the population in $|g; 1/2, -1/2\rangle$ ($\rho_{-}$) remains unchanged because its excitation is suppressed by the presence of the control beam. Half of a period ($T_{p}/2$) later, the precession swaps the populations of $|g; 1/2, +1/2\rangle$ and $|g; 1/2, -1/2\rangle$ states, and “Probe 1-” is fired to measure the original $\rho_{-}$ prior to swapping. The probe pulses are repeated, each with a $T_{p}/2$ delay from the previous pulse [Fig.~\ref{sequence}(a)]. The Bloch vector $S_{z}$ can be calculated as 
 \begin{equation}
\label{eqn00}
S_{z}=\rho_{+}-\rho_{-}=\frac{N_{+}-N_{-}}{N_{+}+N_{-}},
  \end{equation}
 where $\rho_{+}+\rho_{-}=1$, $N_{+}$ and $N_{-}$ are the number of atoms in $|g;\,1/2,\,+1/2\rangle$ and $|g;\,1/2,\,-1/2\rangle$ derived from absorption images of the probe pulses. For each pulse, “Probe 1+” or “Probe 1-”, an absorption image taken by the CMOS camera is compared to a background image taken without atoms to derive an optical depth value at each pixel. Both $\rho_{+}$ and $\rho_{-}$ are measured under the same set of probe conditions, with only $T_{p}/2$ apart in timing. In this way of calculating $S_{z}$, many common-mode imperfections in the laser and detector parameters are suppressed.

For comparison, we have also conducted phase measurements based on the non-QND optical pumping method. Here, the normalization, polarization and probe all use the same laser tuned to the resonance of $6s^2\,{}^1S_0\ (F=1/2) \leftrightarrow 6s6p\,{}^1P_1\ (F=1/2)$ at 399 nm, and no control beam is needed. A different timing sequence is used [Fig.~\ref{sequence}(c)]. A 0.2 ms normalization pulse [“Norm.” in Fig.~\ref{sequence}(c)], fired at $T_{p}/2$ after the initial polarization pulse (“Pol. 1”), measures the total population $N_{+}+N_{-}$. Afterwards the atoms need to be repolarized with “Pol. 2”. Following the free precession time, a 0.2 ms “Probe” pulse ($I/I_{sat}=3\times10^{-4}$) is applied to measures $\rho_{-}$. In this non-QND approach, the probe causes spin flips and can only be applied briefly before the spin information is lost. 

The Bloch vector evolves as $S_{z}=P_{z}\cos(2\pi f t+\phi_{0})$, where $P_{z}$ is the degree of spin polarization, $f$ is the Larmor precession frequency and $\phi_{0}$ is the initial phase. The sinusoidal precession signal, shown in Fig.~\ref{sequence}(d) and \ref{sequence}(e) are obtained by measuring $S_{z}$ at different precession times around 1 s. The measurement uncertainties in the QND approach are significantly reduced in comparison with those in non-QND approach. A key requirement of the QND approach is that the Bloch vector can be repeatedly measured. As shown in Fig.~\ref{sequence}(f), $S_{z,1}$, measured by “Probe 1”, is highly correlated with $S_{z,2}$, measured by “Probe 2”.

\section{\label{efficiency}Measurement Uncertainty}

The Larmor precession phase is determined in the $S_{z}$ measurements, with the highest sensitivity occurring at the points of $\rho_{+}=\rho_{-}$. In the QND approach, the variance of $S_{z}$ can be expressed as (see Appendix \ref{photonshotnoiseQND}), 
\begin{equation}
\label{eqn01}
\sigma^2_{S_{z}}=\sigma^2_{\textbf{op}}+\frac{1}{N_{a}}\simeq\frac{4\overline{p}}{(N_{a}\bar{n})^{2}\epsilon}+\frac{1}{N_{a}},
  \end{equation}
where the first term, denoted $\sigma^2_{\textbf{op}}$, describes the optical noise arising from the fluctuations of both the incoming and the absorbed photons. The second term is the atomic quantum projection noise. All variables in Eq.~(\ref{eqn01}) are defined in Table \ref{tab:table1}. Results of Eq.~(\ref{eqn01}) are approximation for weak absorption cases when $N_{a}\bar{n}/\overline{p}\ll1$. The numerical factor “4” is the total number of images used in the measurement sequence combining “Probe 1+” and “Probe 1-”, with each containing both the absorption and background images. The atomic quantum projection noise is induced by the detection pulse “Probe 1+”, after which all other detection pulses do not contribute any more quantum projection noise. Detailed derivation for both QND and non-QND cases are given in Appendix \ref{photonshotnoise}.

In principle, the excitation cycle in the QND measurement can be repeated indefinitely until the optical noise becomes negligible compared to $1/N_{a}$. In actual experiments, however, the number of excitation cycles is limited by atom losses due to either heating or imperfection in the near-cycling transition. Consider a pulse of $\bar{p}$ photons shot on the atomic clouds through a region of interest of area $A_{\mathrm{ROI}}$, it induces an average number of excitation cycles ($\bar{n}$). The two quantities are related as
\begin{equation}
\label{eqn02}
\exp(-b_{l}\overline{p}\frac{A_{\mathrm{abs}}}{A_{\mathrm{ROI}}})+\bar{n}b_{l}=1,
  \end{equation}
where the first term characterizes the probability for the atom to survive in the bright state, and the second term for the atom to either escape the trap or decay into a dark state. The variables in the equation above are defined in Table \ref{tab:table1}.
The average number of excitation cycles ($\bar{n}$) is affected by the loss branching ratio ($b_{l}$) that takes into account both loss mechanisms. In the QND measurement of this study, the atoms are lost from the trap after an average of 80 excitation cycles due to heating ($b_{l}=1/80$). For comparison, in the non-QND measurement, the spin state is demolished after an average of 3 excitation cycles due to optical pumping ($b_{l}=1/3$). This large difference is indeed the essence of the QND advantage.

\begin{table*}
\caption{\label{tab:table1}
}
\begin{ruledtabular}
\begin{tabular}{cccccccc}
Term & Description & \makecell[c]{Value in the\\ QND approach} & \makecell[c]{Value in the \\ non-QND approach} \\
\hline
$A_{\mathrm{abs}}$& Absorption cross section of an atom & \footnotemark[1] $0.074\ {\rm \mu m}^{2}$ & $0.076\ {\rm \mu m}^{2}$ \\
$A_{\mathrm{ROI}}$& Area of region of interest(ROI) & $8900\ {\rm \mu m}^{2}$ & $8900\ {\rm \mu m}^{2}$  \\
$\overline{p}$& \makecell[c]{Average number of probe photons inside ROI} & $3.4\times10^{6}$ & $0.57\times10^{6}$ \\
$\overline{n}$& \makecell[c]{Average number of excitation cycles} & 23 &2.5 \\
$b_{l}$& \makecell[c]{Loss branching ratio}& 1/80 & 1/3  \\
$N_{a}$ & Number of atoms & $5\times10^{4}$ & $5\times10^{4}$ \\
$\epsilon$ & Quantum efficiency of the camera  & 0.8 (at 556 nm)& 0.4 (at 399 nm)\\
\end{tabular}
\end{ruledtabular}
\footnotetext[1]{The natural absorption cross section of the 556-nm probe transition on resonance is $0.147\ {\rm \mu m}^{2}$. However, the residual Doppler broadening results in a reduced cross section.}
\end{table*}

\begin{figure}
  \includegraphics[width=3.4in]{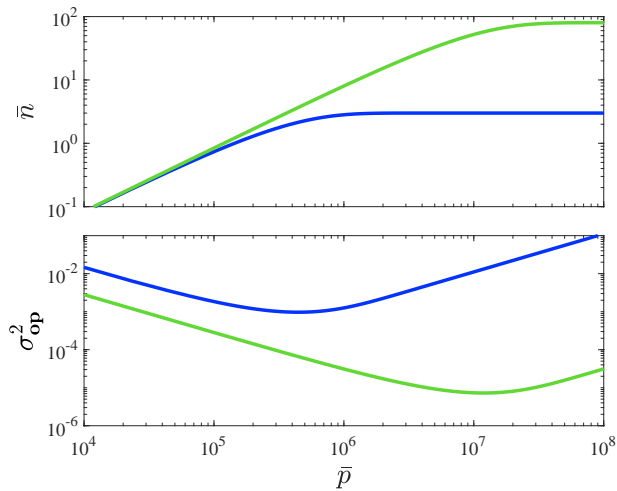}
  \caption{(a) Number of excitation cycles $\bar{n}$ vs. photon number $\overline{p}$ in the probe pulse. In the non-QND case, $\bar{n}$ is limited by spin flipping; In QND, $\bar{n}$ is limited by atom loss from the trap due to heating.
(b)Optical noise $\sigma^2_{\textbf{op}}$ vs. $\overline{p}$. Number of atoms probed is $N_{a}=5\times 10^{4}$. The optimum conditions for QND and non-QND cases occur at different $\overline{p}$ values.
}
\label{var_photon}
\end{figure}

\begin{figure}
  \includegraphics[width=3.4in]{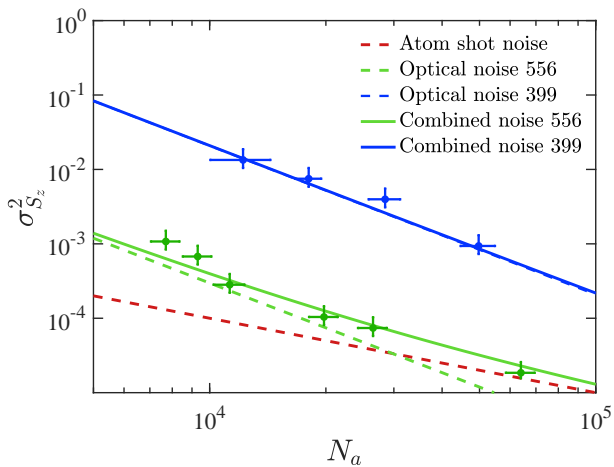}
  \caption{The variance $\sigma^2_{S_{z}}$ vs. the number of trapped atoms $N_{a}$. Blue data points are for non-QND cases, and the green data points for QND cases. The red dashed line indicates the $1/N_{a}$ atomic quantum projection noise. For QND cases, the green dashed line models the optical noise, and the green solid line is combined variance of both the optical noise and the atomic quantum projection noise. The blue dashed line and blue solid line are overlapped.}
\label{efficiency}
\end{figure}

Fig.~\ref{var_photon} shows the average number of excitation cycles $\bar{n}$ and the optical noise $\sigma^2_{\textbf{op}}$ as a function of the photon number ($\overline{p}$) in the probe pulse. When $\overline{p}$ is small, $\bar{n}$ is proportional to $\overline{p}$, while $\sigma^2_{\textbf{op}}$ is inversely proportional to $\overline{p}$. As $\overline{p}$ increases, loss mechanisms come into effect, $\bar{n}$ becomes saturated and $\sigma^2_{S_{z}}$ increases due to the photon shot noise. The optimum choice for $\overline{p}$ occurs at the point when $\bar{n}$ starts to saturate, and $\bar{n}$ is different between QND and non-QND cases. 

Fig.~\ref{efficiency} shows the variances of $S_{z}$ for both the QND and non-QND cases with the number of trapped atoms varying in the range of $10^{3}-10^{4}$. The measured results agree well with calculated ones. From non-QND to QND cases, the optical noise is reduced by 19 dB, independent of the number of atoms. The QND optical noise goes below the atomic quantum projection noise when $N_{a} > 3 \times 10^4$, and is 2.3 dB below at $N_{a} = 5 \times 10^{4}$.

\section{\label{conclusion}Discussion and Outlook}
In this work, we have demonstrated a QND phase measurement of the spin precession of atoms in an optical dipole trap. The 19 dB gain in the optical noise can be further improved by reducing heating loss of the atoms due to optical probing and scattering loss due to impure laser polarization of the control beam. In the current setup, the atoms are transferred into the ODT of 200 $\mu$K depth at a temperature of 100 $\mu$K, and are heated out of the ODT after an average of 80 excitation cycles. By applying laser cooling in the ODT prior to the measurement sequence, the atom temperature could be lowered down toward the Doppler-cooling limit of $4.4\ \mathrm{\mu K}$, thus increasing the number of excitation cycles $\bar{n}$. Laser cooling would also reduce the Doppler-broadened width of 400 kHz, towards the natural linewidth of 182 kHz, and thus increase the photon absorption cross-section. Furthermore, replacing the traveling wave ODT with an optical lattice would increase the trap depth and reduce heating losses due to the probe beam. The impure laser polarization of control beam causes excitation to the $|c\rangle$ state, followed by decay into the lower-lying P levels. This leakage in the cycling transition can be reduced with better polarization control. All these steps would combine to reduce the loss branching ratio $b_{l}$, increase $\bar{n}$ and further suppress the optical noise.

The use of the QND method introduced in this work can be expanded to a wide range of applications. For example, QND measurements can help improve the search sensitivity of a permanent electric dipole moment (EDM) of atoms~\citep{ParDieKal15,2016PhRvC..94b5501B,Zheng2022}. The recently demonstrated tweezer array of Yb atoms, with their spin states acting as qubits, is an emerging platform for quantum computation~\citep{GorReyDal09,Noguchi2011,Jenkins2021a,Ma2022,Barnes2022}, on which the QND approach would help improve readout fidelity. Suppression of spin-flip shown in this work can also be used to decrease spin noise and increase interrogation time in spin squeezing experiments~\citep{Pezze2018}. Moreover, the QND approach can be employed to implement quantum error correction that requires non-destructive detection of error syndromes~\citep{Weiss2017}, as well as real-time feedback control on atomic spin states~\citep{Cox2016,Schleier-Smith2010b}. We emphasize that the all-optical control allows quick switching and real-time programming~\citep{Ma2020}.

While we have focused on the $(F,F+1,F+1)$ ladder-type system, the approach can be generalized to other configurations, such as a $\Lambda$-type
systems or $(F,F+1,F)$ systems. Moreover, instead of the dissipative readout demonstrated in this work, the optical dressing effect can also be applied to dispersive atom-photon interaction~\citep{Braverman2019}, leading to applications such as measurement-based spin squeezing, or generation of entanglement between distant atomic ensembles for distributed quantum sensing~\citep{RN533}.

\begin{acknowledgments}
We would like to thank D. Sheng, Z.-S. Yuan, Y.-G. Zheng, and Y.-N. Lv for helpful discussions. This work has been supported by the National Natural Science Foundation of China through Grants No. 91636215, No. 12174371, No. 11704368, the Strategic Priority Research Program of the Chinese Academy of Sciences through Grant No. XDB21010200, and Anhui Initiative in Quantum Information Technologies through Grant No. AHY110000. C.-L.Z. was supported by NSFC through Grant No. 11922411.

\end{acknowledgments}

\appendix

\section{\label{photonshotnoise}Measurement uncertainty}

\subsection{\label{photonshotnoiseQND}Optical noise in absorption imaging}


Population is detected by measuring the optical depth ($OD$) of the atomic ensemble. $OD$ is derived from the number of detected photons in the reference image without atoms ($p_{1}$) and that in the absorption image of the atomic ensemble ($p_{2}$), 
\begin{equation*}\label{eqn3}
    OD=\ln({p_{1}}/{p_{2}}).
\end{equation*}

For the number of incident probe photons of $p$, and take into account the quantum efficiency of the camera $\epsilon$, the reference image has 
\begin{equation*}\label{eqn1}
    p_{1}=\sum_{i=1}^{p}d_i,
\end{equation*}
where  $d_i$ is a binary variable indexing whether the $i$-th photon is detected: $d_i\sim B(1,\epsilon)$. The number of photons in the absorption image is 
\begin{equation*}\label{eqn2}
    p_{2}=\sum_{i=1}^{p-\sum_{j=1}^{N_{\pm}}n_j}d_i,
\end{equation*}
where $N_{\pm}$ is the number of atoms in the probed state and $n_j$ is the number of photons scattered (absorbed) by the $j$-th atom.

The expectation values and variances for $p_{1}$ and $p_{2}$ can be expressed as
\begin{subequations}%
\begin{eqnarray*}\label{eqn31}
  \mathrm{E}(p_{1})&=&\sigma_{p_{1}}^{2}=\overline{p}\epsilon,\\
  \mathrm{E}(p_{2})&=&\overline{p}\epsilon-\frac{N_{a}}{2}\bar{n}\epsilon,
  \end{eqnarray*}
\begin{eqnarray*}\label{eqn32}
  \sigma_{p_{2}}^{2}
  &=&\mathrm{E}(p-\sum_{j=1}^{N_{\pm}}n_j)\mathrm{Var}(d_i)+\mathrm{Var}(p-\sum_{j=1}^{N_{\pm}}n_j)[\mathrm{E}(d_i)]^2\\
  &=&\overline{p}\epsilon+N_{a}\overline{n}(2\epsilon^{2}-\epsilon)/2+N_{a}\overline{n}^2\epsilon^2/4.
\end{eqnarray*}
\end{subequations}

When calculating $\sigma_{p_{2}}^{2}$, it is assumed that $n_j$ follows the Poisson distribution. 

In our case, $\overline{p}\gg N_{a}\bar{n}\gg 1$. The expectation value and variance for $OD$ are
\begin{subequations}
\begin{eqnarray*}\label{eqn41}
  \mathrm{E}(OD)
  &\simeq&\frac{N_{a}\bar{n}}{2\overline{p}},\\
  \sigma_{OD}^{2}
  &\simeq&\frac{2}{\overline{p}\epsilon}+\frac{N_{a}\overline{n}^{2}}{4\overline{p}^2}.
\end{eqnarray*}
\end{subequations}
These results follow the more detailed derivations found in Ref.~\cite{Lasner2018,Dissertation2019}. It is worth noting that the optical noise in absorption imaging originates from not only the intrinsic photon shot noise discussed above, but also technical noise. In order to reach the fundamental photon-shot-noise level, a total of 30 images without atoms are taken after the detection pulses for a fringe-removal algorithm~\cite{OckTauSpr10}.

\subsection{\label{photonshotnoiseQND}Variance of measured $S_{z}$}

As the populations of $|g;\frac{1}{2},-\frac{1}{2}\rangle$ and $|g;\frac{1}{2},+\frac{1}{2}\rangle$ states are equal, the expectation value and variance for $\rho_{+}$ are
\begin{subequations}
\begin{eqnarray*}\label{eqn43}
  \mathrm{E}(\rho_{+})&=&\frac{1}{2},\\
  \sigma_{\rho_{+}}^{2}
  &\simeq&\frac{5\overline{p}}{2(N_{a}\bar{n})^{2}\epsilon}+\frac{1}{4N_{a}}.\\
\end{eqnarray*}
\end{subequations}

  For non-QND measurements, the variance of measured $S_{z}$ is
  \begin{eqnarray*}\label{eqn62}
    \sigma_{S_{z}}^{2}
    &=&\mathrm{Var}(2\rho_{+}-1)\\
    &\simeq&\frac{10\overline{p}}{(N_{a}\bar{n})^{2}\epsilon}+\frac{1}{N_{a}}.\\
  \end{eqnarray*}
For QND measurements, the variance of measured $S_{z}$ is
\begin{subequations}
  \begin{eqnarray*}\label{eqn63}
    \sigma_{S_{z}}^{2}&=&\mathrm{Var}(\rho_{+}-\rho_{-})\\
   &\simeq&\frac{4\overline{p}}{(N_{a}\bar{n})^{2}\epsilon}+\frac{1}{N_{a}}.
  \end{eqnarray*}
\end{subequations}
The reduction in variances from non-QND to QND is largely due to the difference in the $\bar{n}$ values, and is independent of $N_{a}$.

\begin{figure}[t]
  \includegraphics[width=3.4in]{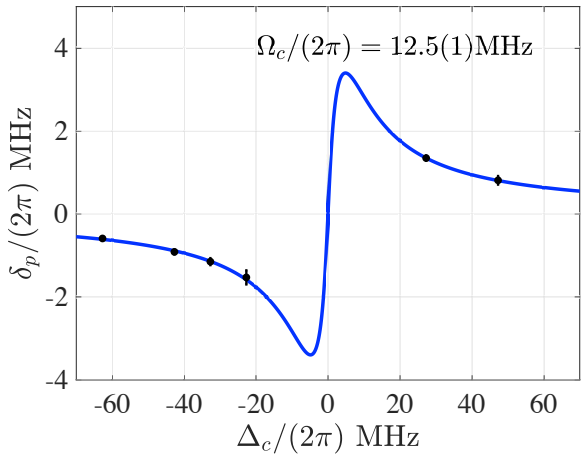}
  \caption{The measured light shift $\delta_{p}$ of the probe transition against the control laser detuning. At a control laser beam of 4 mW and 300 $\mu$m radius, the Rabi frequency $\Omega_{c}/(2\pi)=12.5(1)\,\mathrm{MHz}$ is determined.}
\label{spec423}
\end{figure}

\section{\label{speccontrol}Parameters for the control laser}

The 423 nm control beam drives the $6s6p\,{}^3P_1\leftrightarrow 6s8s\,{}^3S_1$ transition. The natural linewidth of this transition is estimated to be $2\pi \times 1.7$ MHz, based on the known lifetime of the $6s8s\,{}^3S_1$ level\cite{Baumann1985a} and the estimated branching ratios. The detuning $\Delta_{c}$ and the Rabi frequency $\Omega_{c}$ are  experimentally determined by measuring the light shift $\delta_{p}$ of the 556 nm probe transition (Fig.~\ref{spec423}), 
\begin{equation*}
  \delta_{p}= \frac{\Delta_{c}}{2}\ln(1+\frac{2\Omega_{c}^2}{\Gamma_{c}^2+4\Delta_{c}^2}).
\end{equation*}




%

\end{document}